\documentclass[fleqn]{2023SCGE-two}
\usepackage{multirow}
\begin{document}

\ensubject{subject}

%%%%%%%%%%%%%%%%%%%%%%%%%%%%%%%%%%%%%%%%%%%%%%%%%%%%%%%
%%% Authors do not modify the information below
%%% ????????????????
%%% ??????????, ????????????{}, ???????????????????
%Letter to the Editor??Article%??????
\ArticleType{Article}%??Article
\SpecialTopic{SPECIAL TOPIC: }%???????
\Year{}
\Month{}
\Vol{}
\No{}
\DOI{??}
\ArtNo{000000}
\ReceiveDate{}
\AcceptDate{}
%\OnlineDate{January 1, 2016}
%%%%%%%%%%%%%%%%%%%%%%%%%%%%%%%%%%%%%%%%%%%%%%%%%%%%%%%

%%% title: ????
%%%   \title{title}{title for citation}
\title{Identifying the structure of La$_3$Ni$_2$O$_7$ in the pressurized superconducting state}

%%% Corresponding author: ???????
%%%   \author[number]{Full name}{{email@xxx.com}}
%%% General author: ???????
%%%   \author[number]{Full name}{}
\author[1]{Hengyuan Zhang\textsuperscript{\dag}}{}%
\author[1]{Jielong Zhang\textsuperscript{\dag}}{}
\author[1]{Mengwu Huo}{}%\protect\\?��??��????
\author[1]{Junfeng Chen}{}%
\author[1]{Deyuan Hu}{}
\author[1]{Dao-Xin Yao}{}
\author[2]{\\Hualei Sun}{}
\author[1]{Kun Cao}{caok7@mail.sysu.edu.cn}
\author[1]{Meng Wang}{wangmeng5@mail.sysu.edu.cn}
\AuthorMark{Hengyuan Zhang}

\address[1]{Institute of Neutron Science and Technology, Guangdong Provincial Key Laboratory of Magnetoelectric Physics and Devices, School of Physics,\\ Sun Yat-Sen University, Guangzhou, Guangdong 510275, China}
\address[2]{School of Science, Sun Yat-Sen University, Shenzhen, Guangdong 518107, China}

\contributions{These authors contribute equally to this work.}

%%% Author information for page head. ?��?��????????
%%% ??????????????, ??????????author???
\AuthorMark{Zhang H Y}%\authorcr????????

%%% Authors for citation. ????????��????????
%%% ??????????????, ??????????author???
\AuthorCitation{Hengyuan Zhang, Jielong Zhang, Mengwu Huo, Junfeng Chen, Deyuan Hu, Dao-Xin Yao, Hualei Sun, Kun Cao, Meng Wang, Identifying the structure of La$_3$Ni$_2$O$_7$ in the pressurized superconducting state}

%%% Address. ???
%%%   \address[number]{Address, City {\rm Postcode}, Country}

%\contributions{}%????????

%%% Abstract. ??
\abstract{The crystal structure of La$_3$Ni$_2$O$_7$ in its high-pressure superconducting state has been the subject of intense debate, with conflicting reports proposing orthorhombic ($Amam$ or $Fmmm$) and tetragonal ($I4/mmm$) symmetries. Here, using high-pressure Raman spectroscopy down to 3~K, we resolve this controversy by tracking the structural evolution of La$_3$Ni$_2$O$_7$ up to 32.7~GPa. Leveraging rigorous symmetry-based selection rules, we identify a single structural transition from the orthorhombic $Amam$ phase to the $Fmmm$ phase at $\sim$14.5~GPa, signaled by a profound phonon renormalization. Crucially, the persistence of $D_{2h}$ symmetry across the transition rules out the tetragonal $I4/mmm$ phase in the superconducting state in our measurements. The emergence of bulk superconductivity coincides precisely with this transition. Our results establish the orthorhombic $Fmmm$ structure as the intrinsic host of superconductivity in La$_3$Ni$_2$O$_7$ below 19.45 GPa, resolving a central structural controversy and providing a critical foundation for understanding the superconducting mechanism in bilayer nickelates.}%ժҪ

%%% Keywords. ?????


\begin{strip}
\maketitle
\end{strip}
% Restore the SCPMA template's normal body-text leading after the title strip.
\renewcommand{\baselinestretch}{1.1}
\baselineskip 12.02pt
% Restore the SCPMA template's paragraph indentation after the title strip.
\setlength{\parindent}{10.8pt}

%\tableofcontents%?????

%%%%%%%%%%%%%%%%%%%%%%%%%%%%%%%%%%%%%%%%%%%%%%%%%%%%%%%
%%% The main text. ???????
%???????????????????\cref{fig1}
%\twocolumn\onecolumn
%%%%%%%%%%%%%%%%%%%%%%%%%%%%%%%%%%%%%%%%%%%%%%%%%%%%%%%
\section{Introduction}\label{section1}
The discovery of high-temperature superconductivity in Ruddlesden-Popper (RP) nickelates has established a major new family of unconventional superconductors~\cite{Wang2024nor,Wang2025r}. Superconductivity has been observed in bulk samples of bilayer~\cite{Sun2023,Zhang2024h,Hou2023}, trilayer~\cite{Zhu2024s}, and hybrid structural phases~\cite{Shi2025pr,Huang2025s} under high pressure, as well as in thin-film specimens at ambient pressure~\cite{Ko2025,Zhou2025a}.\Authorfootnote Among these, the bilayer compound La$_3$Ni$_2$O$_7$ and its doped derivatives exhibit the highest superconducting transition temperature, $T_c$~\cite{Li2025a,Zhong2025ev,Qiu2025i}. A key structural feature is the bilayer of NiO$_6$ octahedra linked by shared apical oxygen, which is suggested as the fundamental structural unit for superconductivity in RP nickelates~\cite{luo2023b,yang2024o,liu2024e,qu2024,Lu2024i1,Sakakibara2024p,Zhang2024str,Jiang2025t,Yang2023i,Lechermann2023}. In La$_3$Ni$_2$O$_7$, superconductivity emerges above 14~GPa, forming a distinctive right-triangular-like phase diagram~\cite{Li2025i} that contrasts with the commonly observed dome-shaped phase diagrams of heavy-fermion systems, cuprates, and iron-based superconductors~\cite{Crawford2005,Zhang2015Effects, Massidda2016, Medvedev2009E}. Identifying the crystal structure in the superconducting state is therefore a crucial first step toward elucidating the underlying mechanism. However, high-pressure constraints and the technical challenges associated with diamond anvil cells (DACs) have hindered structural investigations in this regime. Consequently, the crystallography of La$_3$Ni$_2$O$_7$ in its superconducting state remains a subject of ongoing debate.

\begin{figure}[t]
\includegraphics[width=0.9\columnwidth]{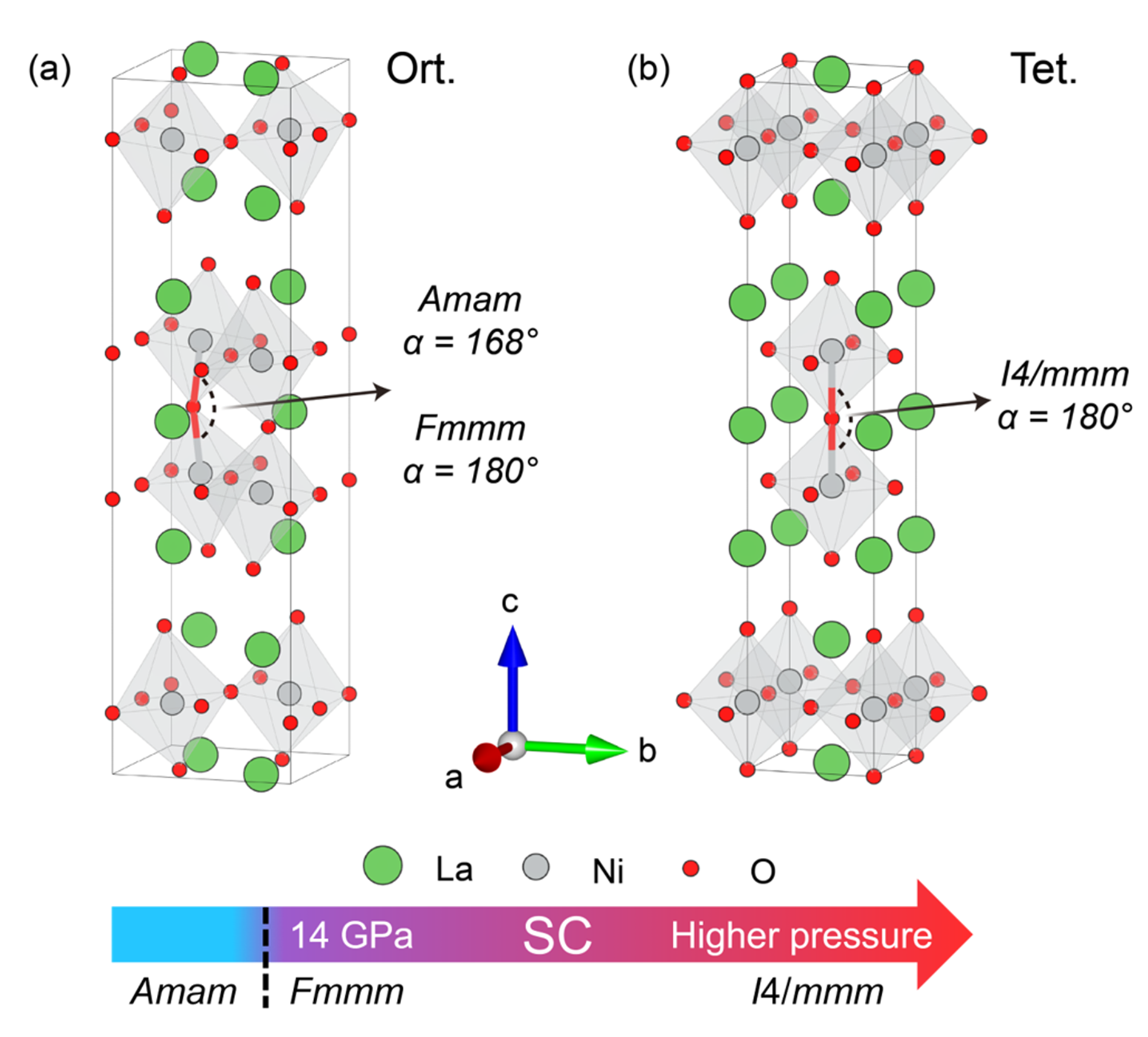}
\centering
\caption{
\textbf{Three candidate crystal structures of La$_3$Ni$_2$O$_7$ under pressure.}
(a) The orthorhombic $Amam$ and $Fmmm$ candidate structures both have $D_{2h}$ point-group symmetry and the lattice constant $a\neq b$  . The $Amam$ structure contains tilted NiO$_6$ octahedra and an interlayer Ni--O--Ni bond angle of approximately 168$^\circ$ along the $c$ axis. In the $Fmmm$ structure, suppression of the octahedral tilt straightens this linkage to 180$^\circ$. (b) The tetragonal $I4/mmm$ candidate structure has $D_{4h}$ point-group symmetry, equivalent in-plane lattice constant $a=b$, and a 180$^\circ$ interlayer Ni--O--Ni bond angle along the $c$ axis. The bottom arrow indicates a possible further $Fmmm$-to-$I4/mmm$ transition at higher pressure. ~\cite{Li2025i}}
\label{fig:1}
\end{figure}

Neutron diffraction measurements at ambient pressure confirm an orthorhombic structure for La$_3$Ni$_2$O$_7$ with a Ni--O--Ni bond angle of 168$^\circ$ along the $c$ axis [Fig. \ref{fig:1}(a)]. However, high-pressure structural studies using synchrotron X-ray diffraction (XRD) have yielded conflicting results, largely because XRD is dominated by scattering from the heavy La and Ni atoms and is relatively insensitive to the positions of light oxygen atoms, while the Ni--O--Ni bond angle is crucial for the structure, and the experimental constraints imposed by the DAC environment are also a challenge. Three primary structural candidates for the superconducting state have been proposed: the unchanged $Amam$ phase~\cite{Shi2025}, an orthorhombic $Fmmm$ phase~\cite{Sun2023,Li2025i}, and a tetragonal $I4/mmm$ phase~\cite{Wang2024s,Wang2025c}, as shown in Figs. \ref{fig:1}(a) and (b). The $Fmmm$ phase, characterized by a 180$^\circ$ Ni--O--Ni bond angle along the $c$ axis, was proposed based on a combination of XRD data at room temperature and first-principles calculations~\cite{Sun2023, Zhang2024eff,Li2025i}. In contrast, the $I4/mmm$ phase was identified from Rietveld refinement of XRD patterns in the superconducting state and the merging of the (0~2~0) and (2~0~0) reflection peaks under pressure~\cite{Wang2025c,Wang2024s}. The reported pressure for this transition varies, either coinciding with or preceding the emergence of superconductivity, and appears sensitive to the pressure-transmitting medium (PTM) and sample form. Consequently, it remains unclear which phase is a prerequisite for superconductivity. As a non-destructive optical method, Raman spectroscopy is inherently sensitive to phonons, making it advantageous for detecting structural symmetry. It has been employed to distinguish the bilayer phase from the other RP nickelates \cite{Li2025d} and investigate the electronic correlations in La$_4$Ni$_3$O$_{10}$ at ambient pressure \cite{suthar2025,Gim2025,Zhang2025d}.

Based on previous high-pressure diffraction studies, we consider the three experimentally proposed candidate structures for pressurized La$_3$Ni$_2$O$_7$: $Amam$, $Fmmm$, and $I4/mmm$. In this work, we employed Raman spectroscopy with polarization analysis to measure the phonon spectra and investigate the structural evolution of La$_3$Ni$_2$O$_7$ under high pressure at 3 and 300 K. Our Raman measurements reveal a clear structural phase transition to a higher-symmetry phase above 14.5 GPa, coinciding precisely with the emergence of pressure-induced superconductivity. Polarized Raman spectroscopy further shows that, already at 1.92 GPa, pressure induces a pronounced evolution of the in-plane Raman-tensor elements of the $A_g$ phonons within the $Amam$ phase. This tensor evolution redistributes phonon spectral weight among the polarization channels without requiring an additional structural transition, although the present data cannot uniquely determine its microscopic origin. Crucially, the $A_g$ phonon modes persist in the $XY$ polarization channel up to 20.48 GPa, thereby providing unambiguous evidence for $D_{2h}$ symmetry in this pressure range and ruling out the tetragonal $I4/mmm$ assignment. Within the candidate-structure framework established by previous high-pressure diffraction studies, the pronounced disappearance of multiple phonon modes at 14.5 GPa reduces the possible high-pressure structure to $Fmmm$ or $I4/mmm$, in accordance with factor-group analysis. Combined with the persistent $D_{2h}$ response in the $XY$ channel, this mode-count reduction identifies the structural transition from $Amam$ to $Fmmm$. By tracking the structural evolution across the pressure--temperature ($P-T$) phase diagram, before and after entering the superconducting state, we identify that the superconducting state of La$_3$Ni$_2$O$_7$ adopts the $Fmmm$ structure below 19.45 GPa. Therefore, the 180$^\circ$ interlayer Ni--O--Ni bond angle, rather than the tetragonal phase, emerges as a key structural prerequisite for high-$T_\text{c}$ superconductivity in bilayer RP nickelates.

 \begin{figure}[t]
  \includegraphics[width=0.9\columnwidth]{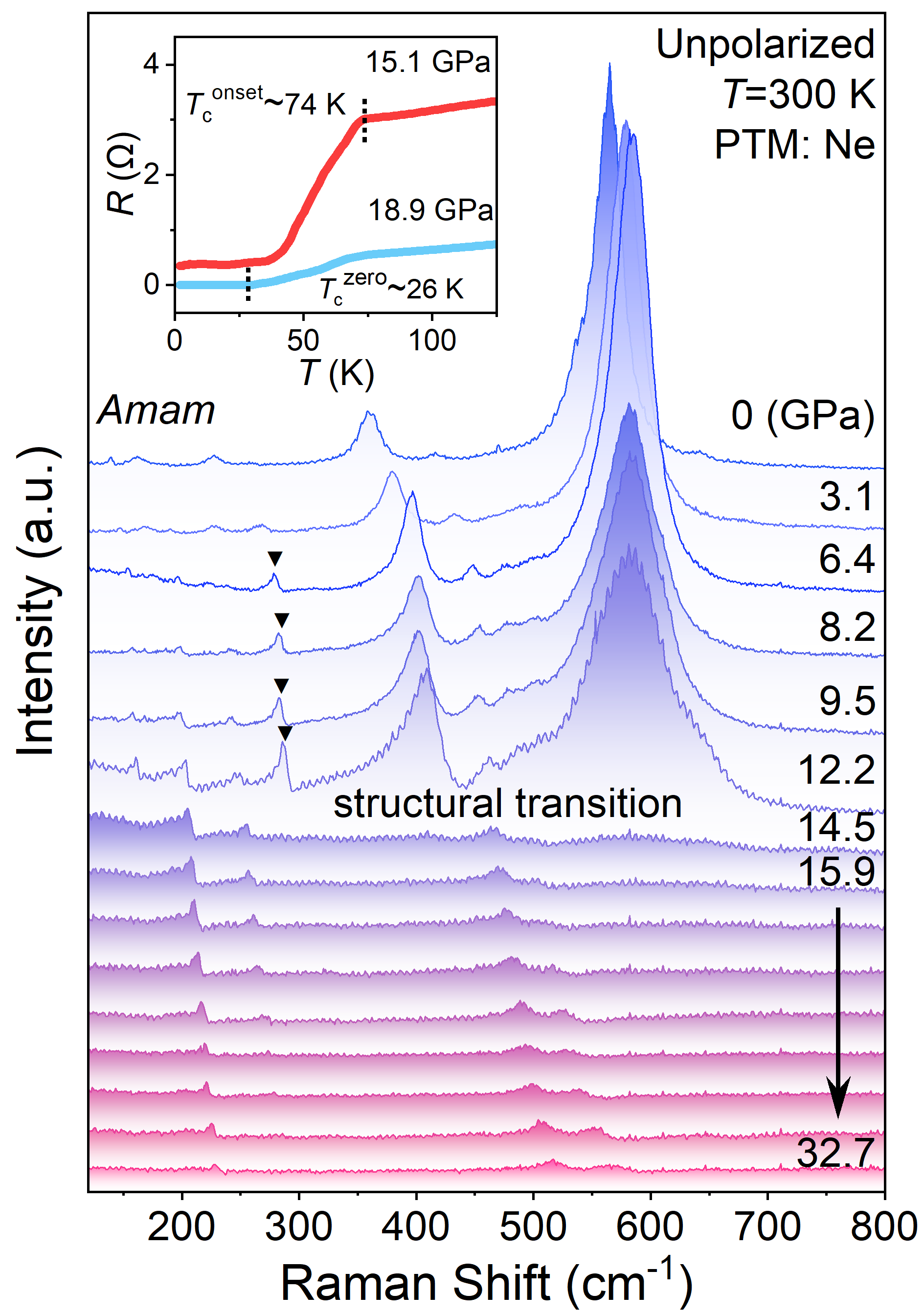}
\caption{\textbf{High-pressure unpolarized Raman spectra of La$_3$Ni$_2$O$_7$.} Raman spectra measured from 0 to 32.7~GPa are vertically offset for clarity. The spectra from ambient pressure to 12.2~GPa are assigned to the orthorhombic $Amam$ phase. Above 14.5~GPa, the phonon spectrum undergoes a pronounced reorganization, signaling a structural transition. All spectra were acquired under identical experimental conditions with a constant sample orientation. The peak marked by $\blacktriangledown$ is associated with an oxygen-octahedral rotational mode and is sensitive to the interlayer Ni--O--Ni bond angle. The inset shows high-pressure resistance measurements, with $T_c^{\mathrm{onset}}=74$~K at 15.1~GPa and $T_c^{\mathrm{zero}}=26$~K at 18.9~GPa. Here, ``unpolarized'' denotes measurements performed without a polarization configuration in the optical path, rather than depolarized Raman scattering.\label{fig:2}}
\end{figure}
\section{Experiments}\label{section2}
Single crystals of La$_3$Ni$_2$O$_7$ were grown using a high-pressure optical floating-zone furnace~\cite{Liu2023e}. Samples from the same batch were previously characterized in high-pressure electrical transport and XRD experiments~\cite{Sun2023,Li2025i,Wang2024s}. The pressure-induced superconductivity was reconfirmed, as shown in the inset of Fig. \ref{fig:2}. For Raman measurements, we cleaved samples along the $ab$-plane and loaded them into a DAC. Neon was used as the PTM for the unpolarized Raman measurements, and He was used as the PTM for polarized Raman measurements. Raman spectra were acquired in a confocal configuration (MoniVista CRS3) using 633 and 532 nm lasers for the unpolarized and polarized measurements, respectively, with the power maintained below 3~mW. We performed low-temperature measurements using an optical magnetic cryostat (Quantum Design), enabling temperature control from 2 to 300~K.  In-situ pressure calibration of the DAC in a low-temperature environment was performed using ruby and diamond. We maintained a constant sample orientation throughout the experiment, ensuring that comparisons of Raman peak intensities are meaningful. 

All Raman scattering measurements were performed in a backscattering geometry, with the incident and scattered light propagating strictly along the crystallographic $c$-axis (i.e., $\mathbf{k}_i \parallel \mathbf{k}_s \parallel [001]$). In this configuration, the polarization vectors of both the incident ($\mathbf{e}_i$) and scattered ($\mathbf{e}_s$) light are confined to the $ab$ plane, determined by Laue diffraction (see Supplementary Fig. S1), such that their $z$-components are strictly zero ($e_{iz} = e_{sz} = 0$). The Raman intensity is governed by $I \propto |\mathbf{e}_s \cdot \mathcal{R} \cdot \mathbf{e}_i|^2$, where $\mathcal{R}$ is the Raman tensor for a specific irreducible representation (see Section E and F in the Supplementary).

\section{Results and Discussions}\label{section3}
Unpolarized Raman spectra of a La$_3$Ni$_2$O$_7$ single crystal measured within the $ab$-plane at pressures up to 32.7~GPa are presented in Fig.~\ref{fig:2}. At ambient pressure, the spectrum of the $Amam$ phase exhibits three primary peaks at 352~cm$^{-1}$, 538~cm$^{-1}$, and 569~cm$^{-1}$ (For further peak resolution, see the Supplementary Fig.~S2), along with several weaker signals. The overall spectral profile and principal peak positions agree well with previous unpolarized Raman spectra measured at ambient pressure~\cite{Wen2024,Li2025d,Meng2024}.  This agreement supports an intrinsic lattice origin for the dominant Raman response of La$_3$Ni$_2$O$_7$ in this work. Oxygen vacancies and other defects may contribute only weakly to the spectra. As the pressure increases to 12.2 GPa, the Raman peaks shift to higher wavenumbers due to lattice compression and broaden progressively, as expected. With increasing pressure, the phonon modes near 284 and 400~cm$^{-1}$ progressively develop asymmetric Fano line shapes, accompanied by slight increases in intensity. We will discuss this phenomenon in the following mode-resolved polarized Raman spectra. Another pronounced transformation occurs at 14.5~GPa, where the spectra show a marked renormalization of phonon modes and the disappearance of the primary peaks at 286, 408, and 600~cm$^{-1}$, clearly indicating a structural phase transition coinciding with the emergence of superconductivity (Fig.~\ref{fig:2}).

Based on the mass-frequency relation $\omega \propto \sqrt{K/M}$, where $K$ is the effective force constant representing the chemical bond stiffness, and $M$ is the effective mass, high-frequency optical phonons ($> 400 \text{ cm}^{-1}$) are predominantly driven by oxygen displacements, whereas lower-frequency vibrations involve the heavier La and Ni atoms. The peak marked by $\blacktriangledown$ is typically associated with oxygen octahedral rotations, which influence the interlayer Ni--O--Ni bond angle (corresponding to 200-300 $\text{cm}^{-1}$), as previously studied in Cu-O or Ni-O perovskite compounds \cite{Gou113, Chaban113, Iliev2006}, while the peaks in the range of 390--520 $\text{cm}^{-1}$ may correspond to Ni--O stretching modes, particularly those involving the apical oxygen \cite{Sugai214, Blumberg1998, AdeAndres_1991, Pintschovius1989, suthar2025}.

\begin{figure}[t]
    \centering
\includegraphics[width=\columnwidth]{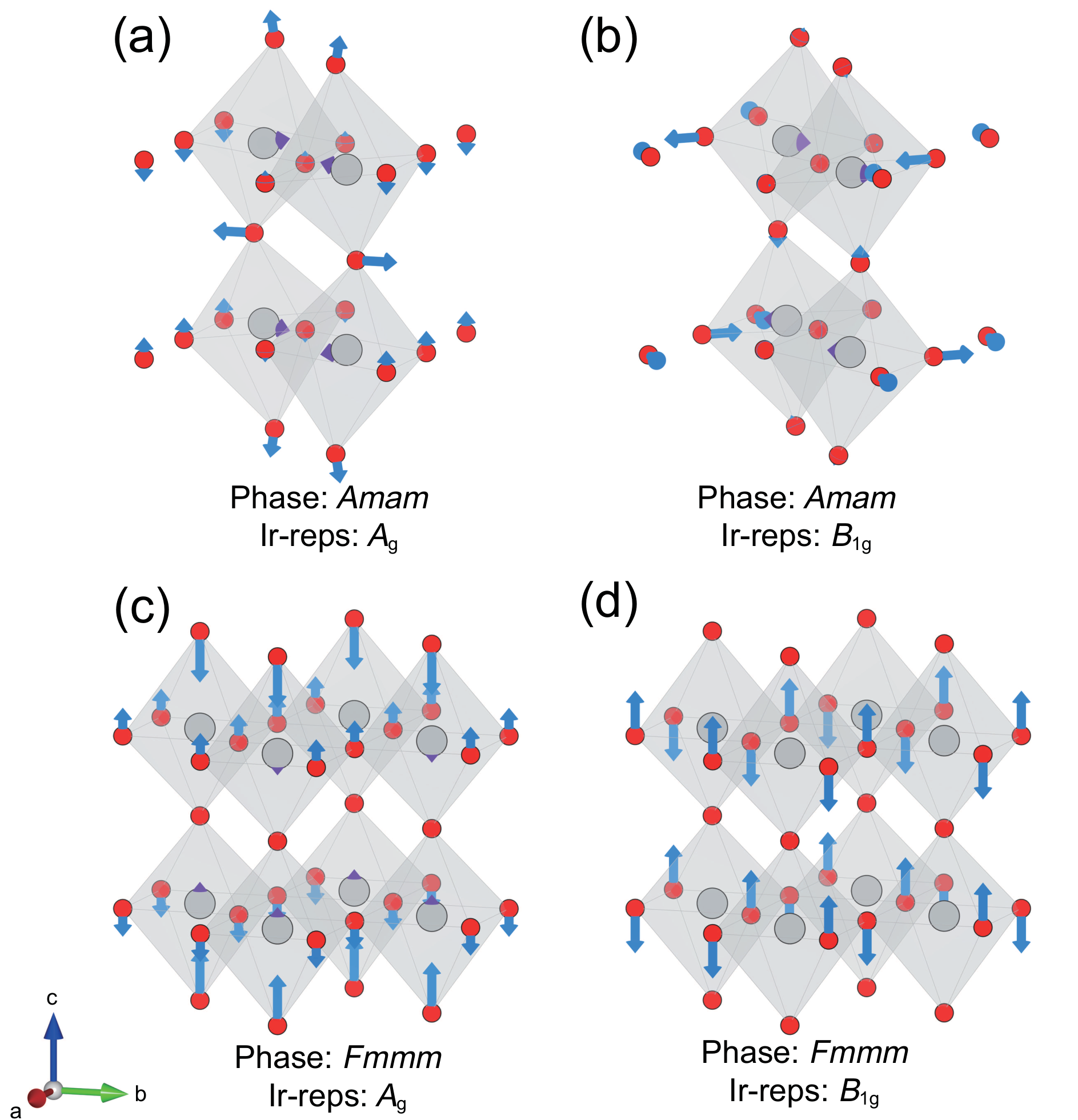}
    \caption{\textbf{Visualization of representative Raman-active vibrational modes under $D_{2h}$ symmetry.} The La atoms are omitted for display, and only a part of the bilayer is represented. (a), (b) The totally symmetric $A_g$ and anti-symmetric $B_{1g}$ modes in the low-pressure $Amam$ phase, where inherent octahedral tilts introduce complex rotational displacement components. (c), (d) The corresponding $A_g$ and $B_{1g}$ modes in the high-pressure $Fmmm$ phase. The pressure-induced straightening of the interlayer Ni--O--Ni bond angle to 180$^\circ$ restores higher local symmetry, driving the observed phonon renormalization at 14.5~GPa.}
    \label{fig:3}
\end{figure}

Factor group analysis at the $\Gamma$ point yields the number of Raman-active modes for the different structural phases (for further details, see Section E and F in the Supplementary). Given our backscattering geometry ($e_{iz} = e_{sz} = 0$), only vibrational modes with nonzero in-plane Raman-tensor elements ($R_{xx}$, $R_{yy}$, $R_{xy}$, or $R_{yx}$) can be experimentally detected, including the $A_g$ and $B_{1g}$ phonon modes in the $Amam$ and $Fmmm$ phases. Furthermore, it is essential to consider whether the specific Wyckoff positions of a given structure give rise to Raman-active modes. Taking into account both our experimental scattering geometry and the crystallographic symmetry constraints, the expected number of observable Raman modes for each phase is given by: $\Gamma_{\text{Raman}}^{Amam} = 10A_g + 12B_{1g}$, 
$\Gamma_{\text{Raman}}^{Fmmm} = 4A_g + 1B_{1g}$, $\Gamma_{\text{Raman}}^{I4/mmm} = 4A_{1g} + 1B_{1g}$. The main phonon modes of oxygen and the corresponding structures are presented in Fig. \ref{fig:3}. It is clear that the active phonon numbers in the Raman spectra are unable to distinguish between the high-pressure $Fmmm$ and $I4/mmm$ phases.

\begin{figure*}[t]
      \centering
      \includegraphics[width=1.8\columnwidth]{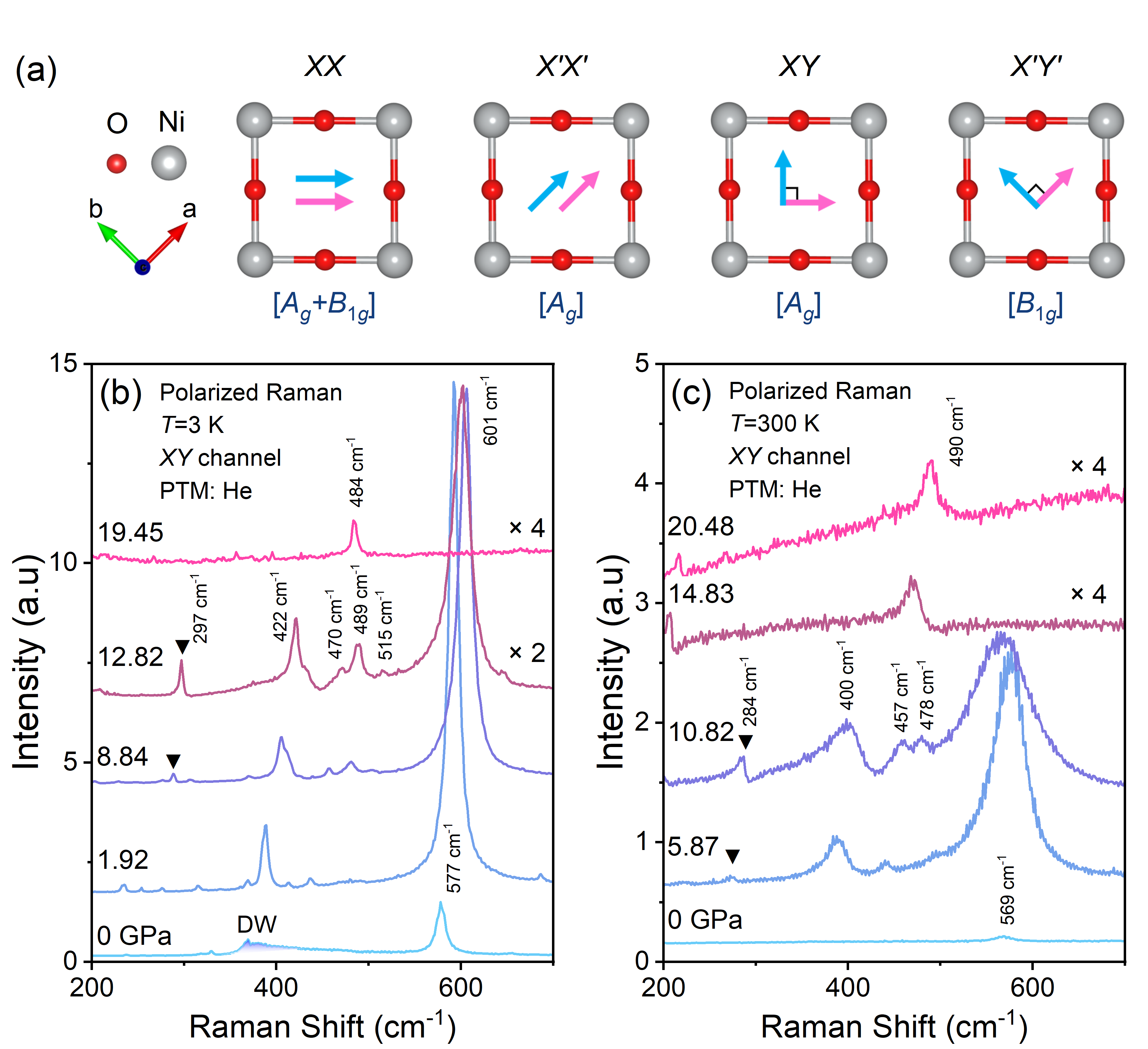}
    \caption{\label{fig:4}\textbf{Polarized Raman spectra of La$_3$Ni$_2$O$_7$.} (a) Schematic of the crystal structure and the definition of the polarization configurations ($XX$, $X^{\prime}X^{\prime}$, $XY$ and $X^{\prime}Y^{\prime}$).  (b) and (c) Pressure-dependent polarized Raman spectra measured in the $XY$ channel at 3 K and 300 K, respectively.}
\end{figure*}

To distinguish the symmetry of the structure in the superconducting state, we performed pressure- and temperature-dependent polarization analyses of the Raman spectra. The measurements were conducted using four standard polarization configurations: $(XX)$, $(X'X')$,$(XY)$ and $(X'Y')$. Here, $X$ and $Y$ represent linear polarizations parallel to the in-plane Ni--O--Ni directions, while $X'$ and $Y'$ are rotated by 45$^\circ$, as shown in Fig. \ref{fig:4}(a).  Table~\ref{Tabs:1} summarizes the phonon modes in various polarization configurations. The most significant difference between the $D_{4h}$ ($I4/mmm$) and $D_{2h}$ ($Amam$ and $Fmmm$) symmetries appears in the $XY$ channel: the former undergoes extinction under ideal polarization conditions, while the latter manifests as an $A_g$ phonon mode. 
\begin{table}[t]
\caption{Raman tensors, calculated intensities, and polarization selection rules for the $Amam$, $Fmmm$, and $I4/mmm$ structural phases. The symbol ``$\times$'' indicates that the corresponding in-plane Raman scattering is forbidden.}
\label{Tabs:1}
\centering
\scriptsize
\renewcommand{\arraystretch}{1.35}
\setlength{\tabcolsep}{1.5pt}
\setlength{\arraycolsep}{1.5pt}

\begin{tabular}{@{}cccccccc@{}}
\toprule
\multirow{2}{*}{\textbf{Phase}} &
\multirow{2}{*}{\shortstack[c]{\textbf{Point}\\\textbf{group}}} &
\multirow{2}{*}{\textbf{Irrep}} &
\multirow{2}{*}{\shortstack[c]{\textbf{Raman}\\\textbf{tensor}}} &
\multicolumn{4}{c}{\textbf{Raman intensity}} \\
\cmidrule(lr){5-8}
& & & & $XX$ & $X'X'$ & $XY$ & $X'Y'$ \\
\midrule

\textit{Amam} &
\multirow{3}{*}{\raisebox{-4.0ex}{$D_{2h}$}} &
$A_g$ &
$\begin{pmatrix}
\alpha & 0 & 0 \\
0 & \beta & 0 \\
0 & 0 & \gamma
\end{pmatrix}$ &
$\frac{(\alpha+\beta)^2}{4}$ &
$\beta^2$ &
$\frac{(\beta-\alpha)^2}{4}$ &
$0$ \\

\textit{Fmmm} &
&
$B_{1g}$ &
$\begin{pmatrix}
0 & \epsilon & 0 \\
\epsilon & 0 & 0 \\
0 & 0 & 0
\end{pmatrix}$ &
$\epsilon^2$ &
$0$ &
$0$ &
$\epsilon^2$ \\
\cmidrule(lr){3-8}

& &
\multicolumn{2}{c}{Selection rules} &
$A_g+B_{1g}$ &
$A_g$ &
$A_g$ &
$B_{1g}$ \\
\midrule

\multirow{3}{*}{\textit{$I4/mmm$}} &
\multirow{3}{*}{\raisebox{-4.0ex}{$D_{4h}$}} &
$A_{1g}$ &
$\begin{pmatrix}
\alpha & 0 & 0 \\
0 & \alpha & 0 \\
0 & 0 & \beta
\end{pmatrix}$ &
$\alpha^2$ &
$\alpha^2$ &
$0$ &
$0$ \\
\addlinespace

& &
$B_{1g}$ &
$\begin{pmatrix}
\gamma & 0 & 0 \\
0 & -\gamma & 0 \\
0 & 0 & 0
\end{pmatrix}$ &
$\gamma^2$ &
$0$ &
$0$ &
$\gamma^2$ \\
\cmidrule(lr){3-8}

& &
\multicolumn{2}{c}{Selection rules} &
$A_{1g}+B_{1g}$ &
$A_{1g}$ &
$\times$ &
$B_{1g}$ \\
\bottomrule
\end{tabular}
\end{table}

Figs.~\ref{fig:4}(b) and (c) present the Raman data in the $XY$ channel at 3 and 300~K, respectively; Supplementary Fig.~S2 shows the data for all four polarization configurations. A small angular offset between the polarizer and analyzer can produce a weak phonon signal in a nominally suppressed channel, namely instrumental polarization leakage; a previous Raman study of La$_3$Ni$_2$O$_7$ reported similar behavior~\cite{He2026Raman}. Therefore, before interpreting the pressure-enhanced $XY$ signal, we first quantified the effective polarization-leakage level. Using the residual intensity of the 577~cm$^{-1}$ $A_g$ phonon in the symmetry-forbidden $X'Y'$ configuration at 0~GPa and 3~K as an internal reference, we estimate an effective polarization-leakage level of approximately 4.4 \% (Supplementary Sec.~G and Fig.~S4). After being pressurized, the $A_g$ phonon intensity in the $XY$ channel is only slightly lower than that in the $X'X'$ channel (where the $A_g$ phonons are strongest). This substantial $XY$ response therefore reflects the intrinsic Raman response of La$_3$Ni$_2$O$_7$, rather than polarization leakage from other channels. In addition, we resolve 14 phonon modes in the complete polarized Raman data set at 0~GPa and 3~K, all of which have counterparts in the independent polarized Raman measurements of He \textit{et al.}~\cite{He2026Raman} (Supplementary Sec.~D and Table~S1). The similar phonon frequencies observed in independent measurements support a predominantly intrinsic lattice origin for the phonons analyzed here rather than oxygen vacancies or other local defects as their dominant origin.

We focus here on the $XY$ spectra in Figs.~\ref{fig:4}(b) and (c). At ambient pressure, the high-frequency $A_g$ phonon considered here has weak intensity in the $XY$ channel. This mode lies near 577~cm$^{-1}$ at 3~K and 569~cm$^{-1}$ at 300~K. At 1.92~GPa and 3~K, the corresponding branch shifts to approximately 592~cm$^{-1}$, and its intensity increases severalfold. As shown in Table~\ref{Tabs:1}, $I_{XY}^{A_g}\propto(\beta-\alpha)^2/4$. For this phonon, the measured polarization intensities yield $\alpha_{0\ GPa}/\beta_{0\ GPa}\approx0.96$ (see Supplementary Sec.~H for details). Therefore, $\alpha\approx\beta$ at ambient pressure, strongly suppressing $I_{XY-0\ GPa
}^{A_g}$, although $D_{2h}$ symmetry allows $A_g$ scattering in the $XY$ channel. 
The quantities $\alpha_\nu=\partial\chi_{xx}/\partial Q_\nu$ and $\beta_\nu=\partial\chi_{yy}/\partial Q_\nu$ are mode-resolved derivatives of the Raman susceptibility with respect to the phonon normal coordinate $Q_\nu$ of the $\nu$-th phonon. Changes in the phonon eigenvector, crystal and electronic structures, electron--phonon coupling, and resonant intermediate states can cause their magnitudes and relative signs to evolve with pressure~\cite{LiuJY,Wen2023}. At 1.92~GPa, $\alpha_{1.92\ GPa}/\beta_{1.92\ GPa}\approx-0.48$ (see Supplementary Sec.~H for details). The opposite signs of $\alpha_{1.92\ GPa}$ and $\beta_{1.92\ GPa}$ therefore enhance the $A_g$ intensity in $XY$ while suppressing it in $XX$ (Supplementary Fig.~S2 and Sec.~H). This spectral-weight redistribution remains fully consistent with the $D_{2h}$ selection rules and does not require a change in point-group symmetry. The present data establish only that pressure changes the in-plane Raman-tensor elements of the $D_{2h}$ $A_g$ phonon, but do not determine the microscopic origin of this evolution. Interestingly, we observe mutually perpendicular stripe-like twin domains in the single crystal, as shown in Supplementary Fig.~S5. Previous studies in YBCO have shown that twinning can modify the measured polarization-dependent intensities\cite{Feile1988,McCartyYBCO,Devereaux2007,Lyzwa2024}, whereas detwinning can restore the polarization-dependent response. However, the quasi-hydrostatic He environment is unlikely to generate sufficient deviatoric stress to induce detwinning. We therefore do not favor pressure-induced detwinning as the origin of the Raman-tensor evolution (see Supplementary Sec.~I for details).

The asymmetry of the phonon modes increases within the pressure range of 1.92--13.78 GPa, as seen in the 284, 400, and 457 cm$^{-1}$ modes at 10.82 GPa and 300 K in Fig.~\ref{fig:4}(c) and Fig.~\ref{fig:2}.  At higher pressure, the system becomes more metallic and exhibits a substantial increase in free charge-carrier density. The interaction between these mobile electrons and localized lattice vibrations is significantly amplified, resulting in enhanced electron--phonon coupling and the characteristic asymmetric Fano lineshapes. This observation is also consistent with the Ni--O--Ni bond angle along the $c$-axis approaching 180$^{\circ}$, where the orbital overlap between the Ni $3d_{z^2}$ and apical oxygen $2p_z$ orbitals becomes more pronounced.

Upon reaching approximately 14.83 GPa, an abrupt phonon renormalization occurs, as shown in Fig.~\ref{fig:4} (c), consistent with the Raman spectra in Fig.~\ref{fig:2}. The sudden disappearance of multiple phonon modes serves as a clear signature of a structural phase transition. Both the $Fmmm$ and $I4/mmm$ phases are expected to exhibit only five observable Raman modes according to factor group analysis. In Fig.~\ref{fig:2}, four phonon modes can be observed. Importantly, one phonon mode at 484~cm$^{-1}$ at 19.45 GPa (3 K) and two phonon modes at 215 and 490~cm$^{-1}$ at 20.48 GPa (300 K) can be observed, as shown in Fig.~\ref{fig:4}. According to polarization analysis, the presence of phonon modes in the $XY$ channel indicates that the system retains its $D_{2h}$ symmetry. In the other polarization channels shown in Fig.~S2, the signal intensity in the $X'X'$ channel is comparable to that in the $XY$ channel. This behavior is consistent with the low-pressure polarization dependence and further supports the retention of $D_{2h}$ symmetry. Thus, the structure in the superconducting state at 19.45 GPa and 3 K is identified as the $Fmmm$ phase. 
The formation of higher structural symmetry causes the octahedral tilting mode (denoted by peak $\blacktriangledown$) to lose its Raman activity above 14.83 GPa. This combination of mode-count reduction and symmetry analysis precisely identifies the phase transition, effectively overcoming the limitations of high-pressure synchrotron XRD in distinguishing $Amam$ from $Fmmm$ because it is inherently insensitive to light oxygen elements.

Having identified the high-pressure structure as the $Fmmm$ phase, we now discuss why some Raman peaks in the unpolarized spectra in Fig.~\ref{fig:2} and polarized spectra in Fig.~\ref{fig:4} persist smoothly across the structural transition without frequency discontinuities. From a symmetry-mode perspective, the structural difference between  $Amam$ and $Fmmm$ can be described by a $Y_2^-$ octahedral-tilt distortion: The $Amam$  structure is obtained primarily by introducing an alternating NiO$_6$-octahedral tilt distortion of the $Fmmm$ lattice, conventionally denoted as the $Y_2^-$ distortion.  Here, $Y_2^-$ denotes an irreducible representation associated with the $Y$ point of the $Fmmm$ Brillouin zone; in the reciprocal-space convention used in Ref.~ \cite{Zhang2024str}, this point is  $Y$ =(0.5,0,0.5).  This distortion bends the apical Ni1--O3--Ni1 linkage away from $180^{\circ}$. Calculations show that its amplitude decreases progressively with pressure and approaches zero near 15~GPa. Because $Y_2^-$ is a zone-boundary distortion, it enlarges the primitive cell relative to $Fmmm$ and reduces the corresponding Brillouin zone. The reduced translational symmetry folds phonons associated with the parent $Y$ point to the $\Gamma$ point of $Amam$, where some become Raman active. When pressure suppresses the $Y_2^-$ distortion, $Fmmm$ recovers its higher translational symmetry and these additional folded modes lose their Raman activity, accounting for the disappearance of many peaks above the transition. In contrast, $\Gamma$-point modes already allowed in $Fmmm$ can persist and evolve continuously under compression. Factor-group analysis shows that $Fmmm$ allows $A_g$ modes involving La1, O1, planar O2, and Ni1, consistent with the persistence of the observed modes involving these atoms. Vibrations involving La1, O1, and planar O2 do not directly correspond to changes in the apical Ni1--O3--Ni1 bond angle and are therefore unlikely to be strongly governed by its straightening. Although Ni1 participates directly in this linkage, the same atom can contribute to different $A_g$ normal modes with distinct displacement directions and relative phases. Suppression of the octahedral tilt therefore does not require these $Fmmm$-allowed modes to disappear or their frequencies to change discontinuously. This is the reason for the smooth evolution of the surviving peaks in Figs.~\ref{fig:2} and~\ref{fig:4} after the structural transition.

A previous powder XRD study proposed an $Amam$ to $I4/mmm$ structural transition near 10~GPa~\cite{Wang2025c}. The reduced transition pressure observed in powder samples may be attributed to a pronounced ``grain clamping effect"~\cite{Guan2018Fra}, whereby the shear stresses arising from numerous grain boundaries overcome the intrinsically weak orthorhombic distortion~\cite{Mandyam2026}. Indeed, high-pressure studies using He as the pressure-transmitting medium can preserve the orthorhombic structure to higher pressures~\cite{Shi2025}.

\section{Summary}\label{section4}
Our polarized high-pressure Raman study identifies a clear structural transition for single crystals of La$_3$Ni$_2$O$_7$ from the $Amam$ phase to the $Fmmm$ phase. At low pressures of approximately 1.92~GPa, compression induces a pronounced change in the in-plane Raman-tensor elements of the $A_g$ phonons. As pressure approaches 14 GPa, the enhanced electron--phonon coupling and characteristic asymmetric Fano lineshapes are consistent with the straightening of the Ni--O--Ni bond along the $c$-axis. At pressures above 14.5~GPa, a pronounced structural phase transition occurs, accompanied by the disappearance of numerous phonon modes. Nevertheless, prominent phonon signals persist in the $XY$ channel, and their polarization dependence resembles that of the low-pressure $Amam$ phase, indicating that the system retains $D_{2h}$ symmetry. Among the three candidate structures proposed for La$_3$Ni$_2$O$_7$, namely $Amam$, $Fmmm$, and $I4/mmm$, the persistence of $D_{2h}$ symmetry rules out the tetragonal $I4/mmm$ structure for the measured sample in the pressure range. Combined with the factor-group analysis of the reduced number of Raman-active modes, these results identify the structural transition near the onset of superconductivity as an $Amam\rightarrow Fmmm$ transition. As in our previous high-pressure XRD studies, the structure of La$_3$Ni$_2$O$_7$ transforms from $Amam$ to $Fmmm$ at 12.3 GPa, and from the orthorhombic to the tetragonal $I4/mmm$ phase at 46.8 GPa. These results reveal that both the $Fmmm$ and $I4/mmm$ phases can host high-temperature superconductivity in bulk La$_3$Ni$_2$O$_7$ under pressure, suggesting that the 180$^\circ$ Ni--O--Ni angle along the $c$-axis is a structural prerequisite for superconductivity.

\Acknowledgements{This work was supported by the National Natural Science Foundation of China (Grants No. 12425404, U25A20193, 92565303, 12494591, 12474137, and 12474249), the Fundamental and Interdisciplinary Disciplines Breakthrough Plan of the Ministry of Education of China (JYB2025XDXM403), the National Key Research and Development Program of China (Grants No. 2023YFA1406000, 2023YFA1406500), the Guangdong Major Project of Basic Research (2025B0303000004), the Guangdong Basic and Applied Basic Research Funds (Grant No. 2024B1515020040, 2025B1515020008, 2024A1515030030), the Shenzhen Science and Technology Program (Grants No. RCYX20231211090245050), the CAS Superconducting Research Project (Grant No. SCZX-0101), the Guangdong Provincial Key Laboratory of Magnetoelectric Physics and Devices (Grant No. 2022B1212010008), and Research Center for Magnetoelectric Physics of Guangdong Province (Grant No. 2024B0303390001)}
\InterestConflict{The authors declare that they have no conflict of interest.}

\bibliographystyle{scpma}
\bibliography{reference}
\end{document}